\documentclass[12pt]{iopart}
\usepackage{epsf}

\newcommand{\Y}{YBa$_2$Cu$_3$O$_{7-\delta}$ }

\begin{document}

\begin{flushright}
\today
\end{flushright}

\article[Oxygen-isotope effect on the in-plane penetration depth
in cuprate superconductors]{}{Oxygen-isotope effect on the
in-plane penetration depth in cuprate superconductors}

\author{R.~Khasanov\dag\ddag\footnote[5]{e-mail:
rustem.khasanov@psi.ch \\ Present address: {\it Laboratory for
Neutron Scattering, ETH Z\"urich and Paul Scherrer Institut,
CH-5232 Villigen PSI, Switzerland; DPMC, Universit\'e de Gen\`eve,
24 Quai Ernest-Ansermet, 1211 Gen\`eve 4, Switzerland;
Physik-Institut der Universit\"{a}t Z\"{u}rich,
Winterthurerstrasse 190, CH-8057 Z\"urich, Switzerland}} ,
A.~Shengelaya\dag, E.~Morenzoni\ddag, K.~Conder\S,
I.~M.~Savi\'c$\|$, and H.~Keller\dag
 }

\address{\dag\ Physik-Institut der Universit\"at Z\"urich, CH-8057
Z\"urich, Switzerland}

\address{\ddag\ Laboratory for Muon Spin Spectroscopy, Paul Scherrer
Institut, CH-5232 Villigen PSI, Switzerland}

\address{\S\ Laboratory for Neutron Scattering, ETH Z\"urich and PSI
Villigen, CH-5232 Villigen PSI, Switzerland}

\address{$\|$\ Faculty of Physics, University of Belgrade, 11001
Belgrade, Serbia and Montenegro}

\begin{abstract}
Muon-spin rotation ($\mu$SR) studies of the oxygen isotope
($^{16}$O/$^{18}$O) effect (OIE) on the in-plane magnetic field
penetration depth $\lambda_{ab}$ in cuprate high-temperature
superconductors (HTS) are presented. First, the doping dependence
of the OIE on the transition temperature $T_c$ in various HTS is
briefly discussed. It is observed that different cuprate families
show a similar doping dependence of the OIE on $T_c$. Then, bulk
$\mu$SR, low-energy $\mu$SR, and magnetization studies of the
total and site-selective OIE on $\lambda_{ab}$ are described in
some detail.  A substantial OIE on $\lambda_{ab}$ was observed in
various cuprate families at all doping levels, suggesting that
cuprate HTS are non-adiabatic superconductors.  The experiments
clearly demonstrate that the total OIE on $T_c$ and $\lambda_{ab}$
arise from the oxygen sites within the superconducting CuO$_2$
planes, demonstrating that the phonon modes involving the movement
of planar oxygen are dominantly coupled to the supercarriers.
Finally, it is shown that the OIE on $T_c$ and $\lambda_{ab}$
exhibit a relation that appears to be generic for different
families of cuprate HTS.  The observation of these unusual isotope
effects implies that lattice effects play an essential role in
cuprate HTS and have to be considered in any realistic model of
high-temperature superconductivity.
\end{abstract}



\maketitle

\section{Introduction}
Although the discovery of the cuprate high-temperature
superconductors (HTS) \cite{Bednorz86} in 1986 triggered
world-wide an enormous effort to understand these novel materials,
there is at present still no convincing microscopic theory
describing the mechanism of superconductivity.  Due to the high
values of the superconducting transition temperature $T_{c}$ and
the early observation of a tiny oxygen-isotope effect in optimally
doped YBa$_2$Cu$_3$O$_{7-\delta}$
\cite{Batlogg87,Batlogg87a,Bourne87}, many theoreticians came to
the conclusion that the electron-phonon interaction cannot be
responsible for high-temperature superconductivity.  As a result,
alternative pairing mechanisms of purely electronic origin were
proposed.  However, the assumption that $T_{c}$ cannot be higher
than 30~K within a phonon-mediated pairing mechanism is not
justified \cite{Carbotte90}.  A prominent example is the recent
discovery of superconductivity in MgB$_{2}$ \cite{Nagamatsu01}
with a $T_{c} \approx 39$~K which is accepted to be a purely
phonon-mediated superconductor.  There is increasing experimental
evidence from recent work, such as neutron scattering
\cite{Mcqueeney99,Chung03}, angle resolved photoemission
spectroscopy \cite{Bogdanov00,Lanzara01}, and isotope effect
studies (this work) that lattice effects play an essential role in
the basic physics of HTS and have to be considered in reliable
theoretical models \cite{Bussmann-Holder01,Eschrig03}.

It is well known that the observation of an isotope effect on
$T_{c}$ in conventional superconductors was crucial in the
development of the microscopic BCS theory.  The isotope shift may
be
quantified in terms of the relation
\begin{equation}
 T_{c} \propto M^{-\alpha} , \ \quad  \alpha = - d\ln T_{c}/d\ln M \, ,
\label{eq:BCS-alpha}
\end{equation}
 where $M$ is the isotope mass, and $\alpha$ is the isotope-effect exponent.  In the
 simplest case of weak-coupling BCS theory $T_{c} \propto M^{-1/2}$ and
 $\alpha_{\rm BCS} \simeq 0.5$, in agreement with a number of experiments on
 conventional metal superconductors.  However, there are exceptions of
 this rule such as Zr and Ru for which $\alpha \simeq 0$.

The conventional phonon-mediated theory is based on the Migdal
adiabatic approximation in which the density of states at the
Fermi level $N(0)$, the electron-phonon coupling constant
$\lambda_{ep}$, and the effective supercarrier mass $m^{\ast}$ are
all independent of the mass $M$ of the lattice atoms.  However, if
the interaction between the carriers and the lattice ions is
strong enough, the Migdal approximation is no more valid
\cite{Alexandrov94}.  Therefore, in contrast to ordinary metals,
unconventional isotope effects on various quantities, such as the
superconducting transition temperature and the magnetic
penetration depth are expected for a non-adiabatic superconductor.

In 1990 the University of Zurich group started a project on
isotope effect on cuprates that was initiated by K.~Alex~M\"uller.
Here we briefly review some of our results. They include
unconventional oxygen isotope ($^{16}$O/$^{18}$O) effects (OIE) in
HTS on the transition temperature and the in-plane magnetic
penetration depth.  For a detailed description of our work we
refer to
Refs.~\cite{Mueller90,Zech95,Zhao98,Mueller00,Zhao01,Keller03,Khasanov03b,
Khasanov03,Khasanov04}. In this review it is demonstrated that the
muon spin rotation ($\mu$SR) technique is a very powerful and
unique tool to investigate the OIE on the magnetic penetration
depth in HTS. In particular, the novel low-energy $\mu$SR
technique (LE$\mu$SR) \cite{Morenzoni97} allows a direct and very
accurate measurement of the magnetic penetration depth in HTS and
with that a reliable measurement of the OIE on $\lambda$.

The paper is organized as follows. In Sec.~\ref{sec:Sample} we
describe the sample preparation and the oxygen exchange procedure.
Some results of the OIE on the transition temperature $T_c$
obtained for different cuprate families are presented in
Sec.~\ref{fig:OIE_Tc}. Sec.~\ref{sec:OIE_lambda} comprises studies
of the OIE on the in-plane penetration depth $\lambda_{ab}$ in
Y$_{1-x}$Pr$_x$Ba$_2$Cu$_{3}$O$_{7-\delta}$ and
La$_{2-x}$Sr$_x$CuO$_4$ by means of bulk $\mu$SR and in optimally
doped YBa$_2$Cu$_{3}$O$_{7-\delta}$ by means of LE$\mu$SR .
Furthermore, results of  the site-selective OIE on $\lambda_{ab}$
in Y$_{0.6}$Pr$_{0.4}$Ba$_2$Cu$_{3}$O$_{7-\delta}$  obtained by
bulk $\mu$SR are reported. In Sec.~\ref{sec:Universal_correaltion}
we discuss implications of the OIE on $\lambda_{ab}$ and the
empirical relation between the isotope effect on $T_c$ and
$\lambda_{ab}$ observed for different HTS families. The
conclusions follow in Sec.~\ref{sec:conclusion}.

\section{Sample preparation and oxygen isotope exchange}\label{sec:Sample}

For isotope effect studies it is important to have isotope
substituted samples of the same quality. In the case of
$^{16}$O/$^{18}$O oxygen substituted samples it means that both
samples should have (i) exactly the same oxygen stoichiometry,
(ii) the same oxygen distribution within the sample, and (iii) the
same grain size distribution (in the case of powder samples).
\begin{figure}[htb]
\vspace{0.5cm}
\begin{center}
\epsfxsize = 15cm \epsfbox{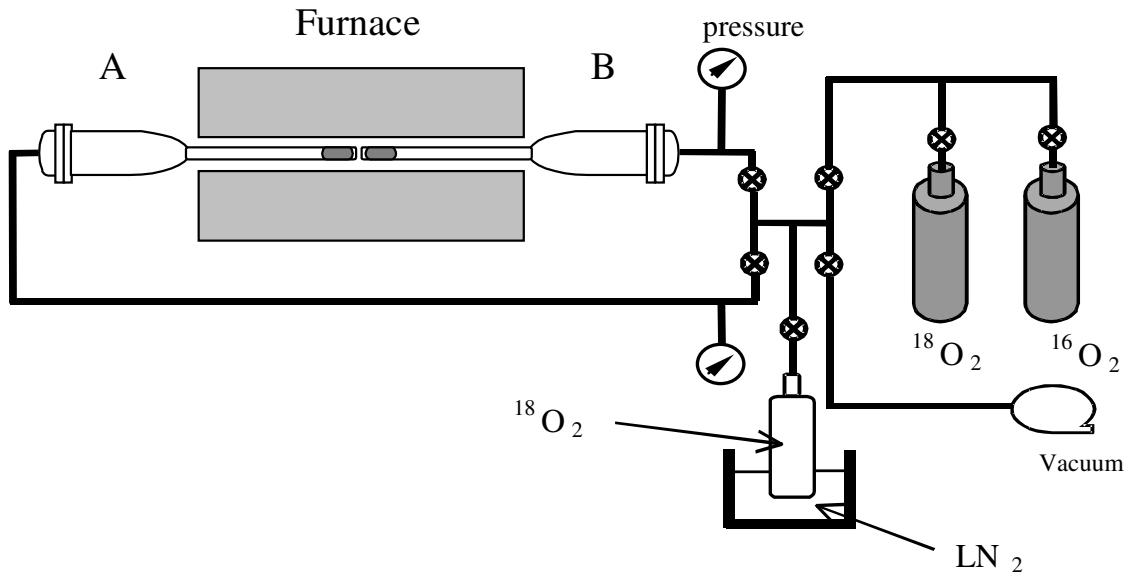}
\end{center}
 \caption{Experimental setup for preparation of the
$^{16}$O/$^{18}$O substituted samples.}
 \label{fig:experimental_setup}
\end{figure}
The schematic view of the experimental setup used for the oxygen
isotope exchange is shown in Fig.~\ref{fig:experimental_setup}
\cite{Conder01}. In order to ensure that the
substituted samples are subject to the same thermal history, the
annealing (in $^{16}$O$_2$ and $^{18}$O$_2$) is performed
simultaneously. In chamber A  the isotope exchange ($^{18}$O) and
in chamber B an identical process in normal oxygen ($^{16}$O)
takes place. A liquid nitrogen trap is used to condense (and
recycle) the expensive $^{18}$O$_2$ after the exchange process is
finished. The exchange apparatus is equipped with a mass
spectrometer (not shown), which allows to view the progress of the
isotope exchange.

\begin{figure}[htb]
 \vspace{0.5cm}
 \begin{center}
\epsfxsize = 13cm \epsfbox{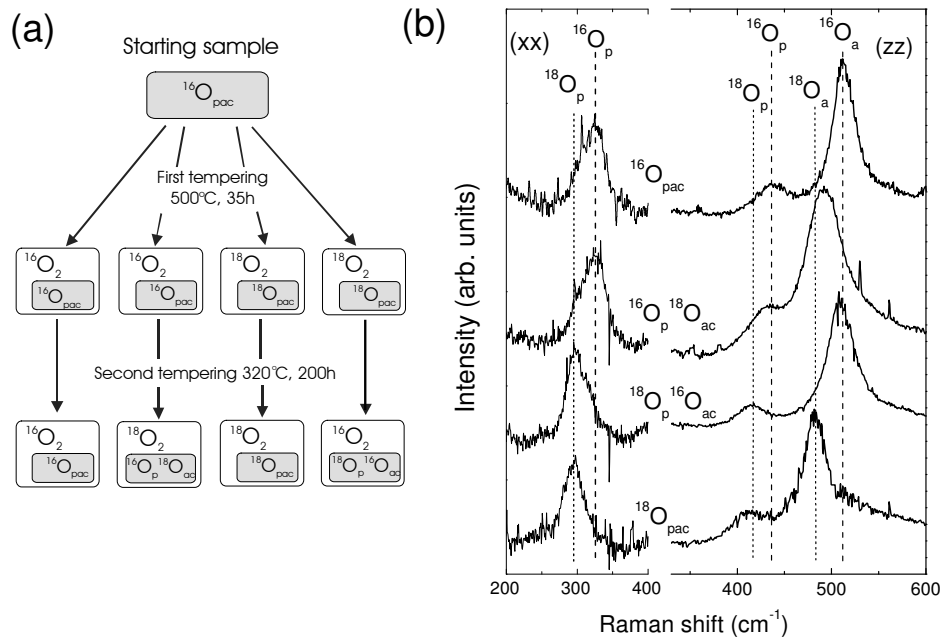}
\end{center}
 \caption{(a) Schematic diagram showing the oxygen
isotope ($^{16}$O/$^{18}$O) substitution procedure to prepare
completely and site-selective substituted samples. The first step
is used to prepare completely oxygen substituted samples. The
second step is used to prepare the site-selective samples from the
completely substituted ones. (b) Room-temperature Raman spectra of
the completely and site-selective oxygen substituted
Y$_{0.6}$Pr$_{0.4}$Ba$_2$Cu$_3$O$_{7-\delta}$ samples. For $xx$
polarization the line corresponds to the out-of-phase motion of
the planar oxygen atoms ($^{16}$O, 325~cm$^{-1}$; $^{18}$O,
294~cm~$^{-1}$). For $zz$ polarization the two lines correspond to
the in-phase motion of the CuO$_2$ plane oxygen ($^{16}$O,
436~$\mathrm{cm}^{-1}$; $^{18}$O, 415~$\mathrm{cm}^{-1}$) and to
the bond-stretching mode of the apical oxygen ($^{16}$O,
512~$\mathrm{cm}^{-1}$; $^{18}$O, 482~$\mathrm{cm}^{-1}$). After
\cite{Khasanov03}. }
 \label{fig:Oxydization}
\end{figure}
The procedure to prepare completely oxygen substituted and
site-selective oxygen substituted
Y$_{1-x}$Pr$_{x}$Ba$_2$Cu$_3$O$_{7-\delta}$ samples (which are
mostly described in this review) is schematically shown in
Fig.~\ref{fig:Oxydization}(a). In the first step [500$^{\rm o}$C,
35~h at 1.2~bar in $^{16}$O$_2$ ($^{18}$O$_2$) gas] completely
oxygen substituted samples ($^{16}$O$_{\rm pac}$ and
$^{18}$O$_{\rm pac}$) are prepared. Here indexes $p$, $a$, and $c$
stay for the planar (within CuO$_2$ planes), the apical and the
chain oxygen, respectively. In order to prepare site-selective
substituted samples after the first step they were grouped in two
pairs. Then two site-selective samples ($^{16}$O$_{\rm
p}$$^{18}$O$_{\rm ac}$ and $^{18}$O$_{\rm p}$$^{16}$O$_{\rm ac}$)
were prepared via annealing one $^{16}$O$_{\rm pac}$ sample in a
$^{18}$O$_2$ atmosphere and one $^{18}$O$_{\rm pac}$ sample in
$^{16}$O$_2$ gas (330$^{\rm o}$C, 150~h, 1.2~bar) [see
Fig.~\ref{fig:Oxydization}(a)]. The other two samples (one
$^{16}$O$_{\rm pac}$ and one $^{18}$O$_{\rm pac}$) were
simultaneously annealed in the same atmosphere as before in order
to have the reference samples following the same thermal history.

The results of the oxygen exchange can be checked by Raman
spectroscopy. Fig.~\ref{fig:Oxydization}(b) shows Raman spectra
with $zz$ and $xx$ polarizations of completely and site-selective
oxygen substituted Y$_{0.6}$Pr$_{0.4}$Ba$_2$Cu$_3$O$_{7-\delta}$
samples \cite{Khasanov03}.
In the $^{18}$O$_{\rm pac}$ sample, the Raman lines are all
shifted to lower frequencies in agreement with the results of Zech
{\em et al.} \cite{Zech96}, indicating a nearly complete exchange
of $^{16}$O with $^{18}$O. In the site-selective sample
$^{16}$O$_{\rm p}$$^{18}$O$_{\rm ac}$, only the position of the
apical oxygen line is shifted to lower frequency, whereas the
lines corresponding to the plane oxygen stay the same [apart from
a small shift of one Raman line (433~cm$^{-1}$ instead of
436~cm$^{-1}$), probably due to a small unintentional partial
substitution by $^{18}$O]. Note that the apical line
(492~cm$^{-1}$) is also slightly shifted from the expected
482~cm$^{-1}$, indicating that the oxygen exchange for the apical
and chain oxygen is slightly uncomplete. In the $^{18}$O$_{\rm
p}$$^{16}$O$_{\rm ac}$ sample only the two planar lines are
shifted, while the apical line stays the same.

\section{Oxygen isotope effect on the transition temperature
$T_c$}\label{sec:OIE_Tc}

\begin{figure}[htb]
\begin{center}
\epsfxsize = 15cm \epsfbox{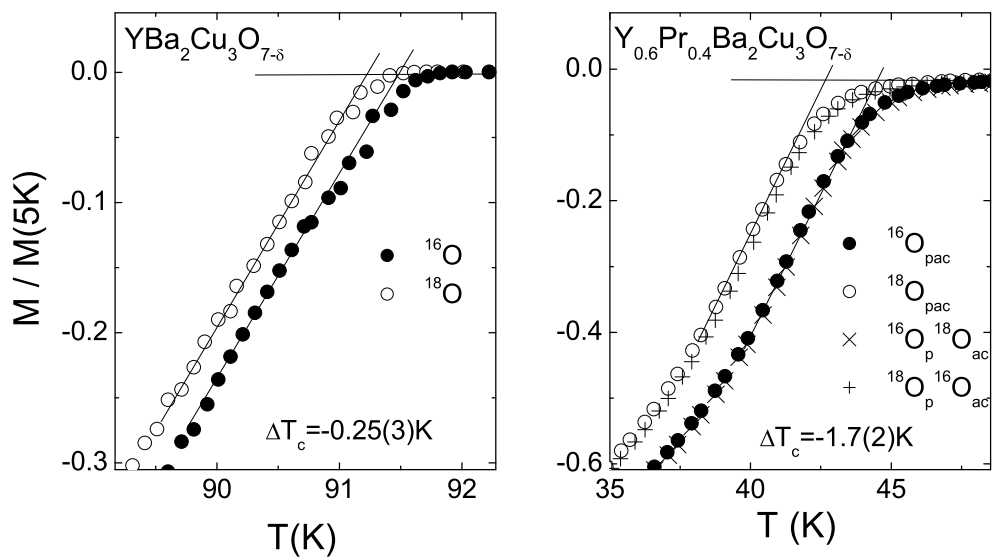}
\end{center}
\vspace{-1cm}
 \caption{Section near $T_c$ of the normalized (to the value at
$5\,{\mathrm{K}}$) magnetization curves (1~mT, FC) of the
completely oxygen substituted YBa$_2$C$_3$O$_{7-\delta}$ (a) and
the site-selective oxygen substituted
Y$_{0.6}$Pr$_{0.4}$Ba$_2$Cu$_3$O$_{7-\delta}$ samples (b). After
\cite{Khasanov03}.}
 \label{fig:OIE_Tc}
\end{figure}
The importance of phonons for the pairing mechanism of
conventional superconductors (including doped fullerenes
\cite{Ramirez92} and MgB$_2$ \cite{Budko01}) was provided by
measurements of the isotope effect (IE) on $T_c$. The BCS theory
predicts that
\begin{equation}
k_B T_c = 1.13 \hbar \omega \exp{\left( -\frac{1}{N(0)V} \right)}
\ ,
 \label{eq:BCS}
\end{equation}
where $\omega$ is a typical phonon frequency (e.g. the Debye
frequency $\omega_D$). The electron phonon coupling
is given by the product of the electron-phonon interaction
constant $V$ and the electronic density of states at the Fermi
surface $N(0)$, both of which are assumed to be independent of the
ion mass $M$. Eq.~(\ref{eq:BCS}) implies an isotope-mass
dependence of $T_c$ ($T_c\propto\omega\propto 1/\sqrt{M}$),
characterized by the isotope effect exponent
$\alpha=1/2$.
%
This is in agreement with isotope effect results reported for many
non-transition metal superconductors (e.g. Hg, Sn and Pd).
However, as mentioned above, there are exceptions as Zr and Ru
with $\alpha\simeq 0$.

Since 1987 a number of oxygen isotope effect investigations on
$T_c$ were performed in most families of HTS. The first OIE
experiments were done on optimally doped samples, showing no
significant isotope shift \cite{Batlogg87,Bourne87}. However later
experiments revealed a small but finite dependence of $T_c$ on the
oxygen isotope mass $M_{\rm O}$
\cite{Batlogg87a,Franck91,Franck94}. It is now well established
that the OIE  on $T_c$ is doping dependent
\cite{Franck94,Schneider92}. As an example, Fig.~\ref{fig:OIE_Tc}
shows the temperature dependence of the magnetization in the
vicinity of $T_c$ in optimally doped $^{16}$O/$^{18}$O
YBa$_2$Cu$_3$O$_{7-\delta}$ samples and site-selective oxygen
exchanged Y$_{0.6}$Pr$_{0.4}$Ba$_2$Cu$_3$O$_{7-\delta}$ samples.
It is seen that the isotope shift on $T_c$ increases with
decreasing doping (increasing Pr concentration) quite
substantially: from $\Delta T_c$=-0.25(3)~K in optimally doped
YBa$_2$Cu$_3$O$_{7-\delta}$ to $\Delta T_c$=-1.7(2)~K in highly
underdoped Y$_{0.6}$Pr$_{0.4}$Ba$_2$Cu$_3$O$_{7-\delta}$.

\begin{figure}[htb]
\begin{center}
\epsfxsize = 12cm \epsfbox{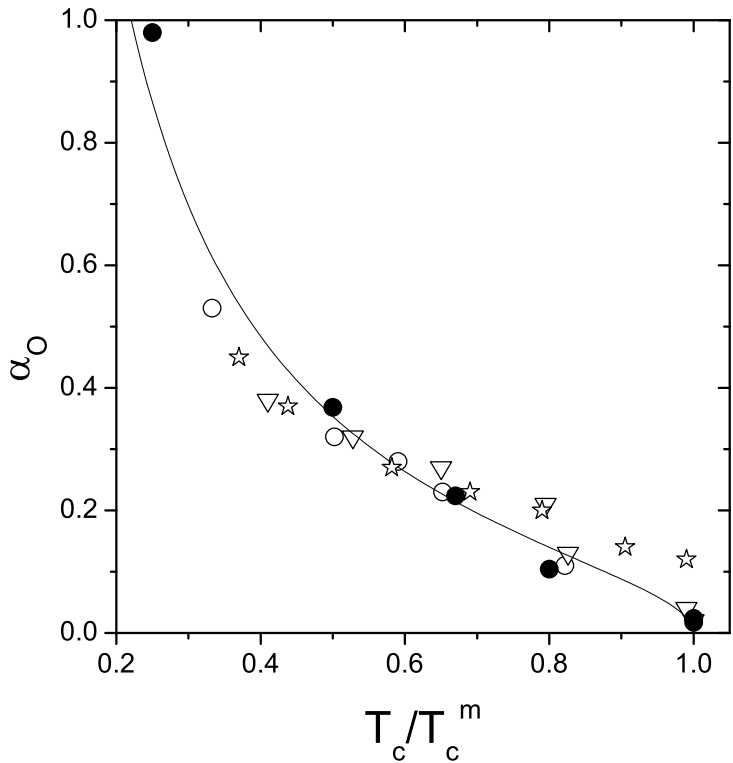}
\end{center}
\vspace{-1cm}
 \caption{Oxygen isotope effect exponent $\alpha_{\rm O}$
versus reduced temperature $\overline{T}_c=T_c/T_c^m$ ($T_c^m$ is
the maximum transition temperature of a given family of HTS). The
open circles are Y$_{1-x}$Pr$_x$Ba$_2$Cu$_3$O$_{7-\delta}$ data
from \cite{Franck91}. Open down triangles are data of
YBa$_{2-x}$La$_x$Cu$_3$O$_7$ taken from \cite{Bornermann91}. Open
stars are data of La$_{1.85}$Sr$_{0.15}$Cu$_{1-x}$Ni$_x$O$_4$ from
\cite{Babushkina91}. Closed circles correspond to
Y$_{1-x}$Pr$_x$Ba$_2$Cu$_3$O$_{7-\delta}$
data from \cite{Khasanov03b,Khasanov04} and unpublished data (this
work). The solid line corresponds to $\alpha_{\rm O} = 0.25
\sqrt{(1-\overline{T}_c)}/\overline{T}_c)$ from
\cite{Schneider92}.}
 \label{fig:alpha_tc}
\end{figure}

The oxygen-isotope exponent  $\alpha_O=-{\rm d}\ln{T_c}/{\rm d}
\ln{M_{\rm O}} $ as a function of doping shows a trend that
appears to be generic for all families of cuprate superconductors
\cite{Zech95,Zhao98,Zhao01,Franck91,Zhao96}. In the underdoped
region $\alpha_O$ is large (even exceeding the BCS value
$\alpha=$1/2) and becomes small in the optimally doped and
overdoped regime (see Fig.~\ref{fig:alpha_tc}).
Moreover, the copper-isotope ($^{63}$Cu/$^{65}$Cu) exponent
$\alpha_{\rm Cu}$ shows a similar trend as $\alpha_{\rm O}$
\cite{Franck93,Zhao96a,Morris98,Williams00}. Both exponents
increase monotonically with decreasing doping level (or $T_c$).
For Y$_{1-x}$Pr$_x$Ba$_2$Cu$_3$O$_{7-\delta}$ and
Y$_{1-x}$Pr$_x$Ba$_2$Cu$_4$O$_{8}$ close to optimal doping
$\alpha_{\rm Cu} \simeq \alpha_{\rm O}$, whereas in the deeply
underdoped regime $\alpha_{\rm Cu} \simeq 0.7 \alpha_{\rm O}$
\cite{Zhao01,Zhao96a,Morris98,Williams00}.

The observation of OIE on $T_c$ indicates that lattice vibrations
in the CuO$_2$ planes are relevant for the occurrence of
superconductivity. From this point of view, one would expect that
the planar oxygen modes give rise to larger OIE on $T_c$ than the
apical and/or chain oxygen modes. The first reliable
site-selective OIE (SOIE) studies of $T_c$ were performed by Zech
{\em et al.} \cite{Zech96}. It was shown that in optimally doped
YBa$_2$Cu$_3$O$_{7-\delta}$ the planar oxygen mainly ($\geq
80$~\%) contribute to the total OIE on $T_c$. These results were
confirmed by Zhao {\em et al.} \cite{Zhao96} and later by Khasanov
{\em et al.} \cite{Khasanov03}. Both groups
\cite{Zhao96,Khasanov03}  found that in underdoped
Y$_{1-x}$Pr$_{x}$Ba$_2$Cu$_3$O$_{7-\delta}$ the predominant
contribution (100~\% within error bar) arises from the oxygen
within the superconducting CuO$_2$ planes [see
Fig.~\ref{fig:OIE_Tc}(b)].

\section{Oxygen isotope effect on the in-plane magnetic
penetration depth $\lambda_{ab}$}\label{sec:OIE_lambda}

The conventional phonon-mediated theory of superconductivity
(standard BCS theory) is based on the Migdal adiabatic
approximation in which the effective supercarrier mass $m^{\ast}$
is independent of the mass $M$ of the lattice atoms.  However, if
the interaction between the carriers and the lattice is strong
enough, the Migdal adiabatic approximation breaks down and
$m^{\ast}$ depends on $M$ (see, {\it e.g.} \cite{Alexandrov94}).
A significant experiment to explore a possible coupling of the
supercarriers to the lattice is an isotope effect study of the
magnetic field penetration depth $\lambda$.

For a general Fermi surface the zero-temperature magnetic
penetration depth $\lambda_{ii}(0)$ may be written as an integral
over the Fermi surface \cite{Chandrasekhar93}:
\begin{equation}
\frac{1}{\lambda_{ii}(0)^{2}} = \frac{\mu_{0}}{2\pi^{2} h} \oint
dS_{F} \frac{v_{F_{i}} \, v_{F_{i}}}{|v_{F}|} \, ,
\label{eq:Fermi-integral}
\end{equation}
where $i$ denotes the crystallographic axes ($a, b, c$) and
$v_{F_{i}}$ is the Fermi velocity.  For the special cases of
spherical or ellipsoidal Fermi surfaces,
Eq.~\ref{eq:Fermi-integral} leads to the London expression
\begin{equation}
\frac{1}{\lambda_{ii}(0)^{2}} = \mu_{0} e^{2}
\frac{n_{s}}{m^{\ast}_{ii}} \, ,
 \label{eq:Fermi-London}
\end{equation}
where $n_{s}$ is the superconducting carrier density and
$m^{\ast}_{ii}$ is the effective mass of the carriers.  For a
general Fermi surface, it is convenient to {\em parameterize}
experimental data by means of Eq.~\ref{eq:Fermi-London}. It should
be noted, however, that in this case $m^{\ast}_{ii}$
is not directly related
the band mass, except for spherical or ellipsoidal Fermi surfaces
\cite{Chandrasekhar93}. For HTS, which are superconductors in the
clean limit, we may parameterize the in-plane penetration depth
$\lambda_{ab}$ in terms of the relation: %
\begin{equation}
1/\lambda_{ab}^{2}(0) \sim n_{s}/m^{\ast}_{ab} \; ,
 \label{eq:Lambda}
\end{equation}
where $m^{\ast}_{ab}$ is the in-plane effective mass (not band
mass) of the charge carriers.  According to Eq.~\ref{eq:Lambda},
this implies that an OIE on $\lambda_{ab}$ has to be due to a
shift in $n_{s}$ and/or $m_{ab}^{\ast}$:
\begin{equation}
\Delta\lambda^{-2}_{ab}(0)/\lambda^{-2}_{ab}(0)= \Delta n_s/n_s
-\Delta m_{ab}^{\ast}/m_{ab}^{\ast}. \label{eq:Deltalambda}
\end{equation}
Therefore a possible mass dependence of $m_{ab}^{\ast}$ can be
tested by investigating the isotope effect on $\lambda_{ab}$,
provided that the contribution of $n_{s}$ to the total isotope
shift is known.

The first observation of a possible OIE on the magnetic
penetration depth $\lambda(0)$ in polycristalline
YBa$_2$Cu$_3$O$_{6.94}$ was reported by Zhao and Morris
\cite{Zhao95}. In another study Zhao {\em et al.}
\cite{Zhao97n,Zhao98} extracted the isotope dependence on
$\lambda(0)$ in fine grained samples of La$_{2-x}$Sr$_x$CuO$_4$
($0.06\leq x\leq 0.15$) from the Meissner fraction. Hofer {\em et
al.} \cite{Hofer00} investigated the OIE on $\lambda^{-2}_{ab}(0)$
in tiny single crystals of La$_{2-x}$Sr$_x$CuO$_4$ by means of
torque magnetometery. All these experiments showed indeed a
pronounced oxygen isotope dependence of the magnetic penetration
depth $\lambda$.

The $\mu$SR technique is a very powerful method to determine the
magnetic penetration depth in superconductors. In the following we
will discuss in more detail OIE investigations of $\lambda_{ab}$
in optimally doped and underdoped cuprate HTS by means of bulk
$\mu$SR and LE$\mu$SR, which at present is the most powerful and
elegant technique to measure the magnetic penetration depth
directly.
Detailed bulk $\mu$SR investigations of polycrystalline samples of
HTS have demonstrated that $\lambda$ can be obtained from the
muon-spin depolarization rate $\sigma(T) \sim 1/\lambda^{2}(T)$,
which probes the second moment of the magnetic field distribution
in the mixed state \cite{Zimmermann95}. It was shown
\cite{Gunn88,Smilga91} that in polycrystalline samples of highly
anisotropic systems such as the HTS ($\lambda_{c}/
\lambda_{ab}>5$), $\lambda_{\rm eff}$ (powder awerage) is
dominated by the shorter penetration depth $\lambda_{ab}$ due to
the supercurrents flowing in the CuO$_{2}$ planes: $ \sigma(T)
\propto 1/\lambda_{ab}^{2}(T)$.
In LE$\mu$SR experiments spin-polarized low-energy muons are
implanted in the thin film sample at a known depth $z$ beneath the
surface and precess in the local magnetic field $B(z)$.  This
feature allows to measure directly the profile $B(z)$ of the
magnetic field inside the superconducting film in the Meissner
state and to make a model independent determination of $\lambda$
\cite{Jackson00}.
All the OIE $\mu$SR results of $T_c$ and $\lambda_{ab}$ discussed
in this review are summarized in Table~\ref{table:OIEresults}. For
comparison the low-field magnetization data are also included.

\begin{table}[htb]
\caption[~]{\label{table:OIEresults} {\it Summary of the OIE
results.}
} %
\begin{center}
\begin{tabular}{lllcccclll}\\ \hline
\hline
Compound & Method &Sample &$\Delta T_c/ T_c$~~&$\Delta
\lambda_{ab}(0)/\lambda_{ab}(0)$~~ \\
 &&shape&(\%)&(\%)\\
\hline
\Y &LE$\mu$SR&thin film&-0.22(16)&2.8(7) \\
\hline
\Y &magnetization~~&fine powder~~&-0.26(5)&3.0(1.1) \\
&&&-0.28(5)$^{\rm a}$&2.4(1.0)$^{\rm a}$\\
\hline
\Y&bulk $\mu$SR&powder&-0.3(1)&2.6(5)\\
Y$_{0.8}$Pr$_{0.2}$Ba$_2$Cu$_3$O$_{7-\delta}$&&&-1.3(3)&2.4(7)\\
Y$_{0.7}$Pr$_{0.3}$Ba$_2$Cu$_3$O$_{7-\delta}$&&&-2.8(5)&2.5(1.0)\\
Y$_{0.6}$Pr$_{0.4}$Ba$_2$Cu$_3$O$_{7-\delta}$&&&-4.6(6)&4.5(1.0)\\
La$_{1.85}$Sr$_{0.15}$CuO$_4$&&&-1.0(1)&2.2(6)\\
\hline
Y$_{0.6}$Pr$_{0.4}$Ba$_2$Cu$_3$O$_{7-\delta}$&bulk $\mu$SR&site-selective&0.1(4)$^b$&0.9(5)$^b$\\
&&powder&-3.7(4)$^c$&3.1(5)$^c$\\
&&&-3.3(4)$^d$&3.3(4)$^d$\\
 \hline \\
\end{tabular}
\begin{flushleft}
\item[] $^{\rm a}$ {results for the back-exchange $^{16}$O $\to$
$^{18}$O and $^{18}$O $\to$ $^{16}$O samples}
\item[] $^{\rm b}${results for the sample with $^{16}$O at plane
sites and $^{18}$O at apical and chain sites}
\item $^{\rm c}${results for completely $^{18}$O substituted
sample}
\item$^{\rm d}${results for the sample with $^{18}$O at plane
sites and $^{16}$O at apical and chain sites }
\end{flushleft}
\end{center}
\end{table}

\subsection{OIE on $\lambda_{ab}$ in underdoped
region}\label{subsec:OIE_underdoped}

The first measurements of OIE on $\lambda_{ab}(0)$ using bulk
$\mu$SR were performed by Khasanov {\em et al.} \cite{Khasanov03b}
in underdoped Y$_{1-x}$Pr$_{x}$Ba$_2$Cu$_3$O$_{7-\delta}$ ($x=0.3$
and $x=0.4$). The transverse-field $\mu$SR measurements were
performed on the beam-line $\pi M$3 at the Paul Scherrer Institute
(PSI, Switzerland) using low-momentum muons (29 MeV/c). The
samples were field-cooled from far above $T_{c}$ in a magnetic
field of 200~mT. The depolarization rate $\sigma$ was extracted
from the $\mu$SR time spectra using a Gaussian relaxation function
$R(t)\propto \exp (-\sigma^{2} t^{2}/2)$.  Above $T_c$ a small
temperature-independent depolarization rate
$\sigma_{nm}=0.15$~$\mu s^{-1}$ is seen, arising from the nuclear
magnetic moments. Below $T_c$, $\sigma$ increases strongly due to
the flux lattice formation.  An additional sharp increase of
$\sigma(T)$ was observed below 10~K which is due to
antiferromagnetic ordering of Cu(2) moments \cite{Seaman90}.
However, zero-field $\mu$SR experiments indicate no presence of
magnetism above 10~K. Therefore, data points below 10~K were
excluded in the analysis. The superconducting contribution
$\sigma_{sc}$ was then determined by subtracting $\sigma_{nm}$
measured above $T_{c}$ from $\sigma$.

\begin{figure}[htb]
\begin{center}
\epsfxsize = 15cm \epsfbox{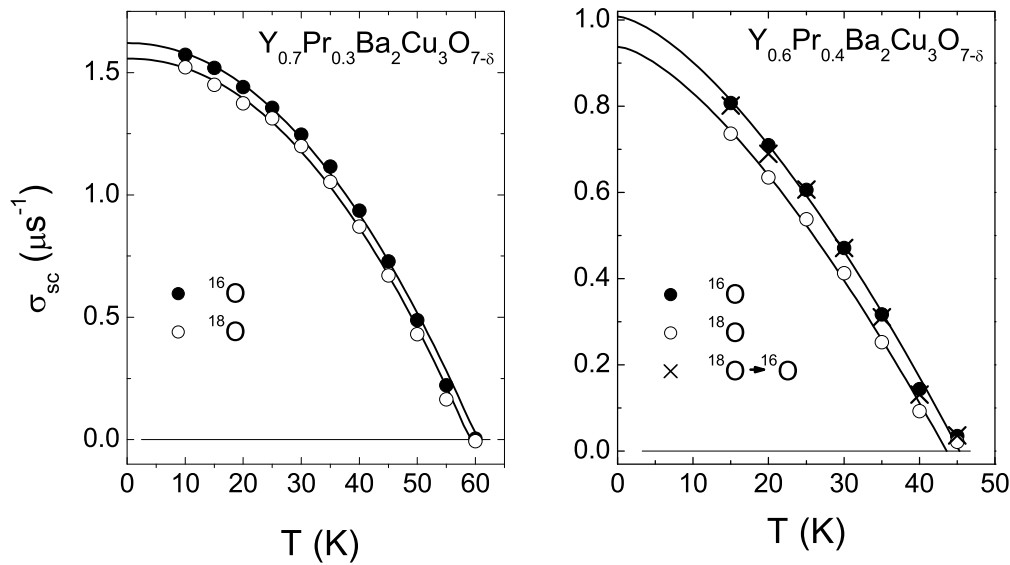}
\end{center}
 \vspace{-1cm}
\caption{Temperature dependence of  the $\mu$SR depolarization
rate $\sigma_{sc}$ of Y$_{1-x}$Pr$_x$Ba$_2$Cu$_3$O$_{7-\delta}$
($x=0.3$ and $x=0.4$) measured in a field 200~mT (FC). The error
bars are smaller than the size of the data points. Data points
below 10~K are not shown (see text for an explanation). The solid
lines correspond to fits to the power law [see
Eq.~(\ref{eq:power_law})]. After \cite{Khasanov03b}.}
 \label{fig:OIE_lambda_YPr_bulk}
\end{figure}
In Fig.~\ref{fig:OIE_lambda_YPr_bulk} the temperature dependence
of $\sigma_{sc}$ for Y$_{1-x}$Pr$_x$Ba$_2$Cu$_3$O$_{7-\delta}$
($x=0.3$ and $x=0.4$) is shown.  It is evident that for both
concentrations $x$ a remarkable oxygen isotope shift on $T_c$ as
well as on $\sigma_{sc}(0)$ is observed.  The data in
Fig.~\ref{fig:OIE_lambda_YPr_bulk} were fitted to the power law
\cite{Zimmermann95}
\begin{equation}
\sigma_{sc}(T)/\sigma_{sc}(0)=1- (T/T_{c})^n.
 \label{eq:power_law}
\end{equation}
The values of $\sigma_{sc}(0)$ obtained from the fits were found
to be in agreement with previous results \cite{Seaman90}. In order
to prove that the observed OIE on $\lambda_{ab}$ is intrinsic, the
$^{18}$O sample with $x=0.4$ was back exchanged
($^{18}$O$\rightarrow ^{16}$O).  As seen in
Fig.~\ref{fig:OIE_lambda_YPr_bulk} the data points of this sample
(crosses) coincide with those of the $^{16}$O sample.  From the
measured values of $\sigma_{sc}(0)$ the relative isotope shift of
$\lambda_{ab}(0)$ was found to be
$\Delta\lambda_{ab}(0)/\lambda_{ab}(0)=2.5(1)\%$ and 4.5(1)\% for
$x=0.3$ and $x=0.4$ samples, respectively (see
Table~\ref{table:OIEresults}).  For the OIE exponent $\beta_{\rm
O}=-d\ln{\lambda_{ab}^{-2}}/d\ln{M_{\rm O}}$ one readily obtains
$\beta_{\rm O}=0.38(12)$ for $x=0.3$ and $\beta_{\rm O}=0.71(14)$
for $x=0.4$.  This means that in underdoped
Y$_{1-x}$Pr$_x$Ba$_2$Cu$_3$O$_{7-\delta}$ the OIE on
$\lambda_{ab}^{-2}(0)$ as well on $T_c$ increases with decreasing
of doping.  This is in excellent agreement with the magnetic
torque results of underdoped La$_{2-x}$Sr$_x$CuO$_4$ single
crystals \cite{Hofer00}.

Having established the existence of OIE on $\lambda_{ab}$, the
following fundamental question arises: Which phonon modes are
responsible for this effect?  Insight in this respect can be
obtained from studying the site-selective OIE (SOIE) on
$\lambda_{ab}$. Khasanov {\em et al.} \cite{Khasanov03} performed
detailed studies of the SOIE in underdoped
Y$_{0.6}$Pr$_{0.4}$Ba$_2$Cu$_3$O$_{7-\delta}$ polycrystalline
samples by bulk $\mu$SR.
 In a series of
experiments $^{16}$O ($^{18}$O) oxygen was selectively substituted
by $^{18}$O ($^{16}$O) at apical ($a$) and chain ($c$) sites,
while $^{16}$O ($^{18}$O) was kept at plane (p) sites, following
the same procedure as described in Refs.~\cite{Zech96,Conder01}.
The site-selectivity of the oxygen exchange was checked by Raman
spectroscopy [see Fig.~\ref{fig:Oxydization}~(b)].
\begin{figure}[htb]
\begin{center}
\epsfxsize = 12cm \epsfbox{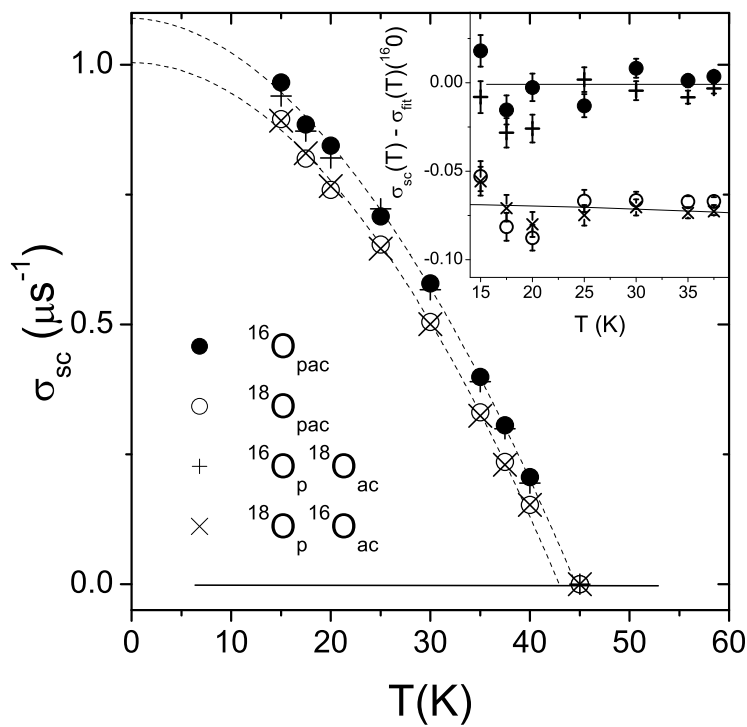}
\end{center}
\vspace{-1cm}
 \caption{ Temperature dependence of the
depolarization rate $\sigma_{sc}$ in site-selective
Y$_{0.6}$Pr$_{0.4}$Ba$_2$Cu$_3$O$_{7-\delta}$ samples  (200~mT,
FC). Data points below 10~K are not shown (see text for an
explanation). The solid lines correspond to fits to the power law
$\sigma_{sc}(T)/\sigma_{sc}(0)=1- (T/T_{c})^n$ for the
$^{16}$O$_{\rm pac}$ and $^{18}$O$_{\rm pac}$ samples. The inset
shows data after subtracting the fitted curve for the
$^{16}$O$_{\rm pac}$ sample.  After \cite{Khasanov03}.}
 \label{fig:SOIE_lambda_bulk}
\end{figure}
Figure \ref{fig:SOIE_lambda_bulk} shows the temperature dependence
of the $\mu$SR depolarization rate $\sigma_{sc}$  for the
Y$_{0.6}$Pr$_{0.4}$Ba$_2$Cu$_3$O$_{7-\delta}$ site-selective
substituted samples.  It is evident that a remarkable oxygen
isotope shift of $T_c$ as well as of $\sigma_{sc}$ is present.
More importantly, the data points of the site-selective
$^{16}$O$_{\rm p}$$^{18}$O$_{\rm ac}$($^{18}$O$_{\rm
p}$$^{16}$O$_{\rm ac}$) samples coincide with those of the
$^{16}$O$_{\rm pac}$($^{18}$O$_{\rm pac}$) samples.
In order to substantiate  these results, the power law curve
[Eq.~(\ref{eq:power_law})] fitting
$\sigma_{sc}(T)$ for $^{16}$O$_{\rm pac}$  was subtracted from the
experimental data (inset in Fig.~\ref{fig:SOIE_lambda_bulk}). It
can be seen that the experimental points for the two pairs of
samples mentioned above coincide within error bar, indicating that
the oxygen site within the CuO$_2$ planes mainly contribute to the
total OIE on $T_c$ and $\lambda_{ab}$.

\subsection{OIE on $\lambda_{ab}$ in optimally doped compounds}
\label{subsec:OIE_optimallydoped}

\begin{figure}[htb]
\begin{center}
\epsfxsize = 12cm \epsfbox{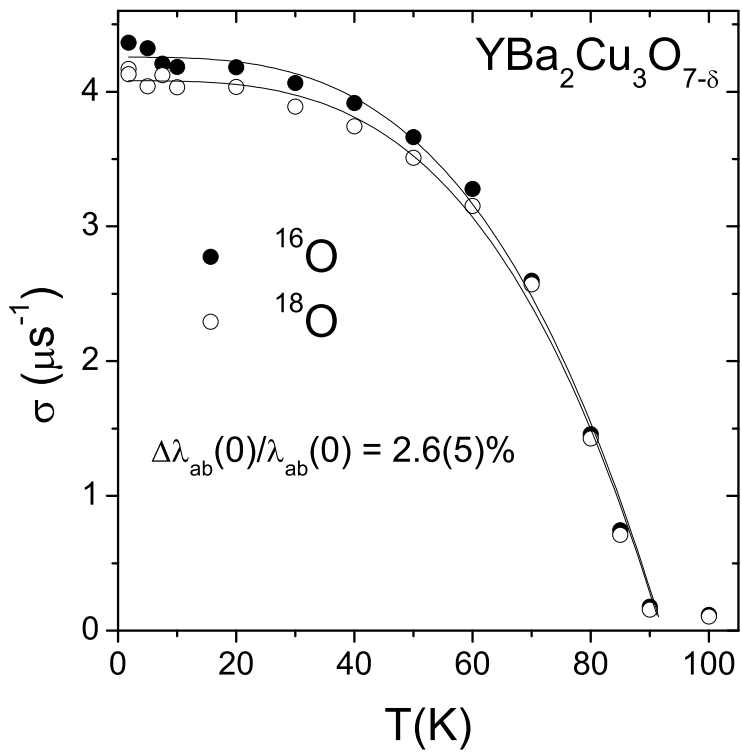}
\end{center}
\vspace{-1cm}
 \caption{Temperature dependence of  the $\mu$SR
depolarization rate $\sigma$ of oxygen isotope substituted
YBa$_2$Cu$_3$O$_{7-\delta}$ measured in a field 200~mT (FC). The
error bars are smaller than the size of the data points. The solid
lines correspond to fits to the power law
[Eq.~(\ref{eq:power_law})].}
 \label{fig:OIE_lambda_bulk}
\end{figure}

Recently, a substantial OIE on $\lambda_{ab}$ in optimally doped
samples of YBa$_2$Cu$_3$O$_{7-\delta}$ [$\Delta\lambda_{ab}(0)/
\lambda_{ab}(0)=2.6(5)$\%, $\beta_{\rm O} = 0.41(7)$] (see
Fig.~\ref{fig:OIE_lambda_bulk}) and La$_{1.85}$Sr$_{0.15}$CuO$_4$
[$\Delta\lambda_{ab}(0)/ \lambda_{ab}(0)=2.2(6)$\%, $\beta_{\rm O}
= 0.35(9)$] was observed by means of bulk $\mu$SR
\cite{KhasanovUnp} (see Table~\ref{table:OIEresults}). The
observation of an OIE on $\lambda_{ab}$ in optimally doped
materials is remarkable considering the small OIE on $T_{c}$
[$\Delta T_{c}/T_{c} = -0.26(5)\%$, $\alpha_{\rm O} \simeq 0.03
$]. The result is also in accordance with magnetization results
obtained for optimally doped
Bi$_{1.6}$Pb$_{0.4}$Sr$_2$Ca$_2$Cu$_3$O$_{10+\delta}$
\cite{Zhao01a}.

\begin{figure}[htb]
\vspace{0.5cm}
\begin{center}
\epsfxsize = 15cm \epsfbox{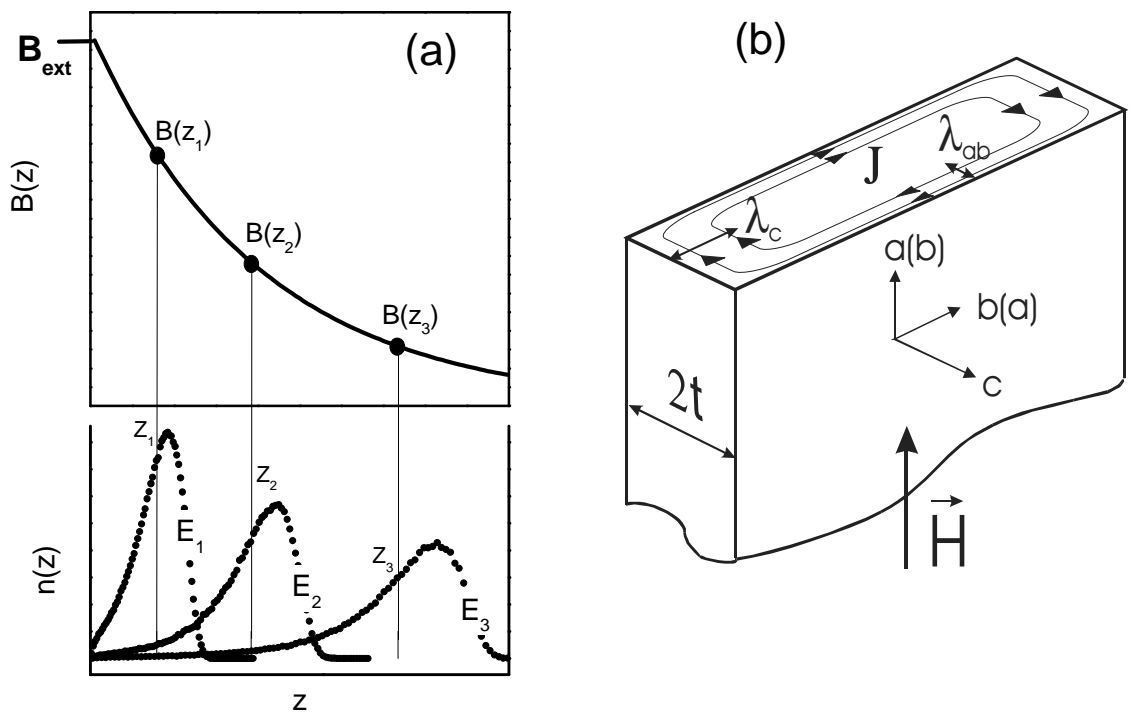}
\end{center}
\vspace{-0.5cm}
 \caption{(a) Principle of $\lambda$ determination
by low-energy $\mu$SR. By tuning the energy of muons they are
implanted at controllable distances ($\overline{z}_1$,
$\overline{z}_2$, or $\overline{z}_3$) from the surface of the
superconductor in a Meissner state. The local magnetic field
$B(\overline{z})$ is determined from the muon precession
frequency.(b) Schematic distribution of the screening current in a
thin anisotropic superconducting slab of thickness $2t$ in a
magnetic field applied parallel to the flat surface. The screening
current $J$ flows preferably parallel to the $ab$ planes, giving
rise to an exponential field decay along the crystal $c$-axis. Due
to twinning in the $ab$ planes ($a$ and $b$ axes are not
distinguishable) the so called in-plane magnetic penetration depth
$\lambda_{ab}$ is measured.
 }
 \label{fig:LEmuSR_principle}
\end{figure}

The most direct and model independent way to confirm this result
is the determination of  $\lambda$ from the functional dependence
of magnetic field profile $B(z)$ penetrating the surface of a
superconductor in the Meissner state. Recently, such a measurement
of $\lambda_{ab}$ was performed in a thin film of
YBa$_2$Cu$_3$O$_{7-\delta}$  at the Paul Scherrer Institute (PSI,
Switzerland) \cite{Jackson00} , by using the novel low-energy
$\mu$SR (LE$\mu$SR) technique \cite{Morenzoni94}. The principle of
the measurement is shown schematically in
Fig.~\ref{fig:LEmuSR_principle}. Polarized muons of energy in the
keV range are implanted into the film. By tuning their energy
these particles can be stopped at different and controllable
depths beneath the surface of the superconductor in the Meissner
state. The stopping profile of the muons, shown in the bottom part
of Fig.~\ref{fig:LEmuSR_principle}(a), depends on the energy and
can be reliably simulated \cite{Morenzoni02,Morenzoni04}. From the
known value of the mean implantation depth $\overline{z}$ and the
corresponding average field  $\overline{B}$ obtained from the
$\mu$SR precession spectra, the functional dependence
$B(\overline{z})$ is determined.

By means of this technique, measurements of OIE on $\lambda_{ab}$
in optimally doped c-axis oriented YBa$_{2}$Cu$_{3}$O$_{7-\delta}$
thin films were recently performed.  A weak external magnetic
field of $9.2$~mT was applied parallel to the sample surface after
the sample was cooled in zero magnetic field from a temperature
above $T_c$ to 4~K. In this geometry (the thickness of the sample
is negligible in comparison with the width), currents flowing in
the $ab$-planes determine the magnetic field profile along the
crystal $c$-axis inside the film [see
Fig.~\ref{fig:LEmuSR_principle}(b)]. Spin-polarized muons were
implanted at depths ranging from 20-150~nm beneath the surface of
the film by varying the energy of the incident muons from 3 to
30~keV.
\begin{figure}[htb]
\begin{center}
\epsfxsize = 12cm \epsfbox{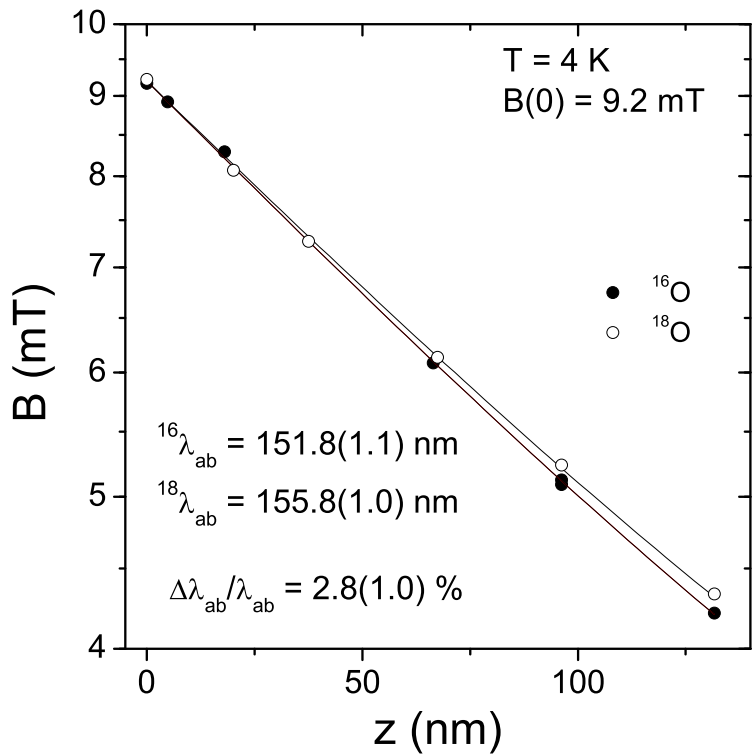}
\end{center}
\vspace{-1cm}
 \caption{Magnetic field penetration profiles $B(z)$
on a logarithmic scale for a $^{16}$O substituted (closed symbols)
and a $^{18}$O substituted (open symbols) \Y film measured in the
Meissner state at 4~K and an external field of 9.2~mT, applied
parallel to the surface of the film.  The data are shown for
implantation energies 3, 6, 10, 16, 22, and 29~keV starting from
the surface of the sample.  Solid curves are best fits by
Eq.~(\ref{eq:exp_decay_film}). Error bars are smaller than the
size of symbols. After \cite{Khasanov04}.
 }
 \label{fig:OIE_lambda_LEMU}
\end{figure}
For each implantation energy the average value of the magnetic
field $\bar B$  and the correspondent average value of the
stopping distance $\bar z$ were extracted.  The value of $\bar B$
was taken from the fit of the time evolution of the muon-spin
polarization spectrum by using the Gaussian relaxation function
and $\bar z$ was the first moment of the simulated $n(z)$
distribution. Results of this analysis for the $^{16}$O and
$^{18}$O substituted \Y films are shown in
Fig.~\ref{fig:OIE_lambda_LEMU}. The magnetic penetration at the
surface of the superconductor is more pronounced for the $^{18}$O
substituted film showing immediately that $^{18}\lambda_{ab} >
^{16}\lambda_{ab}$. The solid lines represent a fit to the $\bar
B$ data by the function:
\begin{equation}
\label{eq:exp_decay_film}
B(z)=B(0)\frac{\cosh[(t-z)/\lambda_{ab}]}{\cosh(t/\lambda_{ab})}
\; .
\end{equation}
This is the form of the classical exponential field decay in the
Meissner state $B(z)=B(0) \exp (-z/\lambda_{ab})$ (where $B(0)$ is
the field at the surface of the superconductor), modified for a
film with thickness $2t$ with flux penetrating from both sides.
The value of $z$ was corrected in order to take into account the
surface roughness of the films \cite{Jackson00}, which was taken
equal ($8.0(5)$~nm) for both samples originating from the same
batch. Fits with Eq.~(\ref{eq:exp_decay_film}) to the extracted
$^{16} \bar B(\bar z)$ and $^{18} \bar B(\bar z)$ yield
$^{16}\lambda_{ab}({\rm 4K})=151.8(1.1)$~nm and
$^{18}\lambda_{ab}({\rm 4K})=155.8(1.0)$~nm. Taking into account a
$^{18}$O content of 95\%, the relative shift is found to be
$\Delta\lambda_{ab}/\lambda_{ab}=(^{18}\lambda_{ab}-^{16}\lambda_{ab})/^
{16}\lambda_{ab} = 2.8(1.0)$\% at 4~K. This value is consistent
with earlier estimates of the OIE on $\lambda$ for optimally doped
\Y \cite{Zhao95,Zhao96}, La$_{1.85}$Sr$_{0.15}$CuO$_4$
\cite{Zhao01a} and
Bi$_{1.6}$Pb$_{0.4}$Sr$_{2}$Ca$_{2}$Cu$_{3}$O$_{10+\delta}$
\cite{Zhao01a} extracted {\it indirectly} from magnetization
measurements.

In order to prove the intrinsic character of the effect, a
backexchange experiment was performed on fine powder samples,
where the isotope dependence of $\lambda_{ab}(0)$ can be
determined from that of the Meissner fraction $f$
\cite{Khasanov04}.
Such type of measurements have been previously performed on the
La$_{2-x}$Sr$_x$CuO$_4$ system in a wide range of doping
($0.06\leq x \leq 0.15$) \cite{Zhao98,Zhao97n}.
\begin{figure}[htb]
\begin{center}
\epsfxsize = 12cm \epsfbox{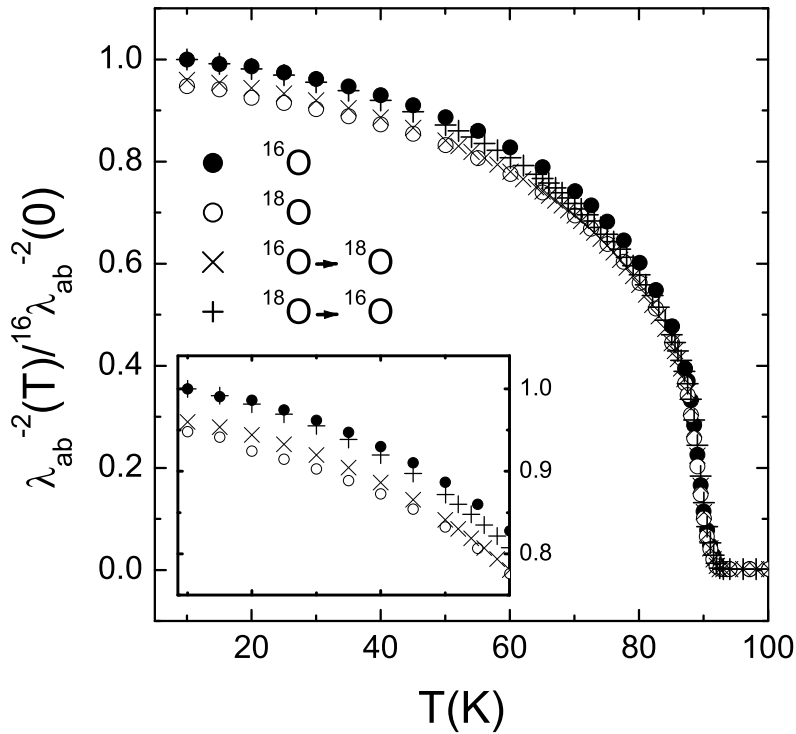}
\end{center}
\vspace{-1cm}
 \caption{Temperature dependence of
$\lambda_{ab}^{-2}$ normalized by $^{16}\lambda_{ab}^{-2}(0)$ for
$^{16}$O and $^{18}$O substituted YBa$_2$Cu$_3$O$_{7-\delta}$ fine
powder samples as obtained from low-field SQUID magnetization
measurements. The inset shows the low-temperature region between
10~K and 60~K. The reproducibility of the oxygen exchange
procedure was checked by the backexchange (crosses). After
\cite{Khasanov04}.
 }
 \label{fig:OIE_lambda_Magn}
\end{figure}
Figure~\ref{fig:OIE_lambda_Magn} shows the temperature dependence
of $\lambda_{ab}^{-2}$ calculated from $f$
for the $^{16}$O/$^{18}$O substituted YBa$_2$Cu$_3$O$_{7-\delta}$
fine powder samples.
The value of $f$ was determined from the FC SQUID magnetization
measurements taken at 1~mT . The absence of weak links between
grains is confirmed by the linear magnetic field dependence of the
FC magnetization measured at 5~K in 0.5~mT, 1~mT, and 1.5~mT.
$\lambda_{ab}(T)$ can be determined from the measured
$\lambda_{\rm eff}(T)$ using the relation $\lambda_{\rm
eff}=1.31\lambda_{ab}$ which holds for highly anisotropic
superconductors ($\lambda_c/\lambda_{ab}>5$)
\cite{Gunn88,Smilga91}. The relative shift at 4~K is found to be
$\Delta\lambda_{ab}/\lambda_{ab}=3.0(1.1)$\%, in good agreement
with the bulk $\mu$SR and LE$\mu$SR data (see
Figs.~\ref{fig:OIE_lambda_bulk}, \ref{fig:OIE_lambda_LEMU} and
Table~\ref{table:OIEresults}). The results of the back exchange
experiments shown as crosses in Fig.~\ref{fig:OIE_lambda_Magn}
prove the intrinsic character of the effect.

\section{Implications of the OIE on $\lambda_{ab}$ and
universal correlations between OIE on $T_c$ and
$\lambda_{ab}$}\label{sec:Universal_correaltion}

From the present investigations it is evident that there is an OIE
on $\lambda_{ab}(0)$ at all doping levels which appears to be
generic for cuprate HTS.  Here the fundamental questions arise: Is
the observation of an OIE on the zero-temperature penetration
depth (superfluid density) a direct signature of strong lattice
effects? And what is the relevance of this finding for the pairing
mechanism? In order to adequately answer these questions more
experimental, and in particular more theoretical, work is needed.
In the following we only discuss a few points related to these
questions.

Under the assumption that we can {\em parameterize} experimental
data of $\lambda_{ab}(0)$ in terms of Eq.~(\ref{eq:Lambda}) it
follows from Eq.~(\ref{eq:Deltalambda}) that the isotope shift of
$\lambda_{ab}$ is due to an isotope shift of $n_s$ and/or
$m^\ast_{ab}$.  It was demonstrated by a number of independent
experiments that for La$_{2-x}$Sr$_x$CuO$_4$
\cite{Zhao98,Zhao01,Zhao97n,Hofer00} and
Y$_{1-x}$Pr$_x$Ba$_2$Cu$_3$O$_{7-\delta}$ \cite{Khasanov03b} the
change of $n_s$ during the oxygen exchange procedure is negligibly
small.  Recently, Khasanov {\em et al.} \cite{Khasanov04} provided
further evidence of this scenario from measurements of the nuclear
quadrupole resonance (NQR) frequency of the plane and the chain
$^{63}$Cu in $^{16}$O and $^{18}$O substituted optimally doped
YBa$_2$Cu$_3$O$_{7-\delta}$ powder samples.  It was found that the
change of the hole number per unit cell caused by the isotope
substitution is less than $\simeq 10^{-3}$.  Since in optimally
doped YBa$_2$Cu$_3$O$_{7-\delta}$ there is approximately one doped
hole per unit cell, the relative change of the hole density
$\Delta n/n$ at the oxygen exchange must be less than $10^{-3}$.
The absence of an observable OIE on $n_s$ implies that the change
of $\lambda_{ab}(0)$ is mainly due to a change on $m^{\ast}_{ab}$.
From Eqs.~(\ref{eq:Lambda}) and (\ref{eq:Deltalambda}) it follows
that
\begin{equation}
\Delta m_{ab}^{\ast}/m_{ab}^{\ast} \simeq -
\Delta\lambda^{-2}_{ab}(0)/\lambda^{-2}_{ab}(0)= 2
\Delta\lambda_{ab}(0)/\lambda_{ab}(0) . \label{eq:Deltamstar}
\end{equation}
This implies that in HTS $m^{\ast}_{ab}$ depends on the oxygen
mass $M_{\rm O}$ with $\Delta m_{ab}^{\ast}/m_{ab}^{\ast} \simeq 5
- 10\%$, depending on doping level (see also
Table~\ref{table:OIEresults}). Note that such an isotope effect on
$m^*_{ab}$ is {\em not expected} for a conventional weak-coupling
phonon-mediated BCS superconductor.  In fact in HTS the charge
carriers are very likely coupled to optical phonons, as indicated
by measurements of the static and high-frequency dielectric
constants \cite{Alexandrov00}, neutron scattering
\cite{Mcqueeney99,Chung03}, photoemission
\cite{Bogdanov00,Lanzara01,Shen01} and tunnelling
\cite{Ponomarev99} experiments.  Strong interaction between the
charge carriers and the lattice ions leads to a {\it break down}
of the adiabatic Migdal approximation \cite{Alexandrov94}, and
consequently the effective supercarrier mass $m^\ast$ depends on
the mass of the lattice atoms. To our knowledge there are just a
few theoretical models which predict an OIE on the effective
carrier mass $m^*$ (see {\it e.g.}
Refs.~\cite{Alexandrov94,Scalapino87,Grimaldi98,Bussmann03}).

\begin{figure}[htb]
\begin{center}
\epsfxsize = 12cm \epsfbox{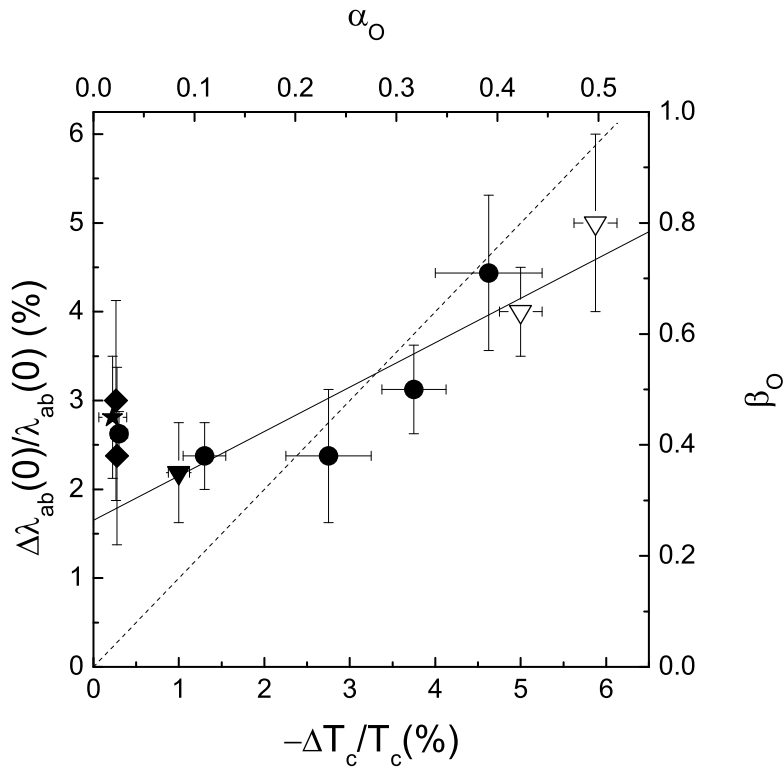}
\end{center}
\vspace{-1cm}
 \caption{\small Plot of the OIE shift $\Delta
 \lambda_{ab}(0)/\lambda_{ab}$(0) (OIE exponent $\beta_{\rm O}$) versus
 the OIE shift $ - \Delta T_c/T_c$ (OIE exponent $\alpha_{\rm O}$) for
 Y$_{1-x}$Pr$_{x}$Ba$_2$Cu$_3$O$_{7-\delta}$ and
 La$_{2-x}$Sr$_{x}$CuO$_{4}$. Closed circles and triangles are bulk $\mu$SR
 data for Y$_{1-x}$Pr$_{x}$Ba$_2$Cu$_3$O$_{7-\delta}$ and
 La$_{1.85}$Sr$_{0.15}$CuO$_4$ (Table~\ref{table:OIEresults}).
 Diamonds and stars are LE$\mu$SR and Meissner fraction data for
 optimally doped YBa$_2$Cu$_3$O$_{7-\delta}$
 (Table~\ref{table:OIEresults}).  Open triangles are torque
 magnetization data for La$_{2-x}$Sr$_x$CuO$_4$ from \cite{Hofer00}.
 The dashed line corresponds to
 $\Delta\lambda_{ab}(0)/\lambda_{ab}(0)=|\Delta T_c/T_c|$.  The solid
 line indicates the flow to 2D-OSI criticality and provides with
 Eq.~(\ref{eq:2D-QSI-criticality}) an estimate for the oxygen isotope
 effect on $d_{s}$, namely $\Delta d_{s}/d_{s} = 3.3(4) \%$.}
 \label{fig:OIE_UnCorr}
\end{figure}

Khasanov {\em et al.} \cite{Khasanov03b} reported an empirical
linear relation between the OIE exponents $\alpha_{\rm O}= -d\ln
T_{c}/d\ln M_{\rm O}$ and $\beta_{\rm O}= - d\ln
\lambda_{ab}^{-2}(0)/d\ln M_{\rm O}$ for underdoped
La$_{2-x}$Sr$_x$CuO$_4$ and
Y$_{1-x}$Pr$_x$Ba$_2$Cu$_3$O$_{7-\delta}$. In
Fig.~\ref{fig:OIE_UnCorr} we plot $\Delta
\lambda_{ab}(0)/\lambda_{ab}(0) \propto \beta_{\rm O}$ versus
$-\Delta T_{c}/T_{c} \propto \alpha_{\rm O}$ for
Y$_{1-x}$Pr$_{x}$Ba$_2$Cu$_3$O$_{7-\delta}$ and
La$_{2-x}$Sr$_{x}$CuO$_{4}$ for the data listed in
Table~\ref{table:OIEresults}.
It is evident from Fig.~\ref{fig:OIE_UnCorr} that there is
correlation between the OIE of $T_{c}$ and $\lambda_{ab}(0)$ which
appears to be universal for cuprate HTS.  Different experimental
techniques (SQUID magnetization, magnetic torque, bulk $\mu$SR,
LE$\mu$SR) and different types of samples (single crystals,
powders, thin films) yield within error bar consisting results,
indicating that the observed {\em isotope effects are intrinsic}
and not artifacts of the particular experimental method or sample
used.  As indicated by the dashed line, at low doping level
$\Delta \lambda_{ab}(0)/\lambda_{ab}(0) \simeq |\Delta
T_{c}/T_{c}|$, whereas close to optimal doping $\Delta
\lambda_{ab}(0)/\lambda_{ab}(0)$ is almost constant and
considerably larger than $|\Delta T_{c}/T_{c}|$ ($\Delta
\lambda_{ab}(0)/\lambda_{ab}(0) \approx 10 |\Delta T_{c}/T_{c}|$).
This finding is consistent with the empirical ''Uemura relation``
\cite{Uemura89,Uemura91} and with the ''parabolic ansatz``
proposed in \cite{Schneider92} in a {\em differential way} for
doped cuprate HTS.

It was shown by Schneider and Keller \cite{Schneider01,Keller04}
that this empirical relation naturally follows from the doping
driven 3D-2D crossover and the 2D quantum superconductor to
insulator (2D-QSI) transition in the highly underdoped limit.
Close to the 2D-QSI transition the following relation holds
\cite{Keller04}:
\begin{equation}
\Delta\lambda_{ab}(0)/\lambda_{ab}(0) = (1/2) [\Delta d_{s}/d_{s}
-  \Delta T_{c}/T_{c}],
 \label{eq:2D-QSI-criticality}
\end{equation}
where $\Delta d_{s}/d_{s}$ is the oxygen-isotope shift of the
thickness of the superconducting sheets  $d_{s}$ of the sample.
The solid line in Fig.~\ref{fig:OIE_UnCorr} represents a best fit
of Eq.~(\ref{eq:2D-QSI-criticality}) to the data with $\Delta
d_{s}/d_{s}= 3.3(4) \%$. Since the lattice parameters are not
modified by oxygen substitution \cite{Conder94,Raffa98}, the
observation of an isotope shift of $d_s$ implies local lattice
distortions involving oxygen which are coupled to the superfluid
\cite{Keller04}.
\begin{figure}[htb]
\begin{center}
\epsfxsize = 12cm \epsfbox{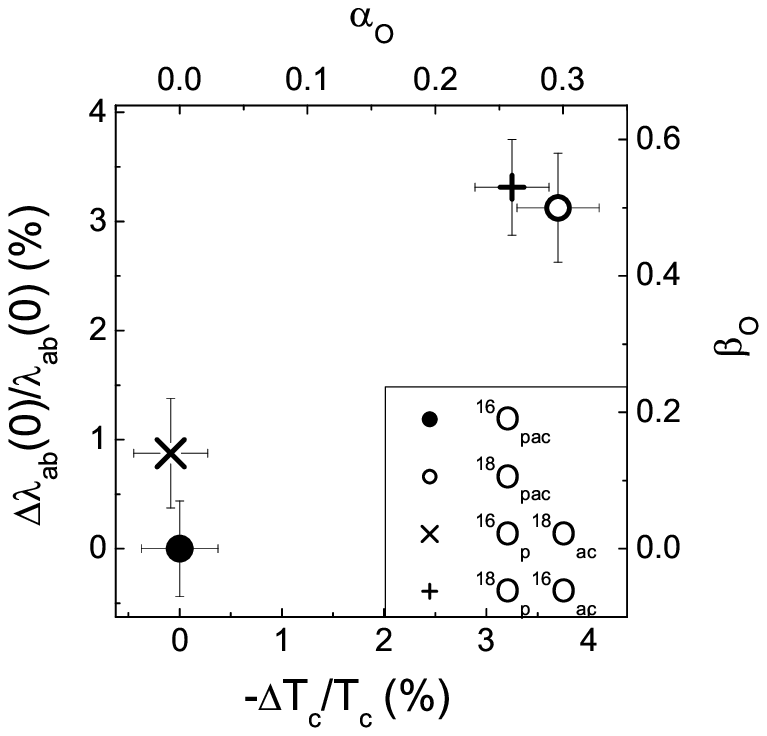}
\end{center}
\vspace{-1cm} \caption{\small Plot of the  OIE  shift of the
in-plane penetration depth $\Delta
\lambda_{ab}(0)/\lambda_{ab}(0)$ (OIE exponent $\beta_{\rm O}$)
versus the OIE shift of the transition temperature -$\Delta
T_c/T_c$ (OIE exponent $\alpha_{\rm O}$) for site-selective
substituted Y$_{0.6}$Pr$_{0.4}$Ba$_2$Cu$_3$O$_{7-\delta}$ samples.
The error bars of the  ``trivial'' $^{16}$O$_{\rm pac}$ point
(zero isotope shift by definition) indicate the intrinsic
uncertainty in the determination of $T_c$ and $\lambda_{ab}$.
After \cite{Khasanov03}.}
\label{fig:SOIE_UnCorr}%
\end{figure}

As shown in Fig.~\ref{fig:SOIE_UnCorr}, a similar type of
correlation between the OIE on $\lambda_{ab}(0)$ and OIE on
$T_{c}$ is also observed for the site-selective substituted
Y$_{0.6}$Pr$_{0.4}$Ba$_2$Cu$_3$O$_{7-\delta}$ samples
\cite{Khasanov03}.  It is evident that the oxygen site within the
CuO$_2$ planes mainly contributes to the total OIE on $T_c$ and
$\lambda_{ab}(0)$.  Since for fixed Pr concentration the lattice
parameters remain essentially unaffected by the isotope
substitution \cite{Conder94,Raffa98}, these results unambiguously
demonstrate the existence of a coupling of the charge carriers to
phonon modes involving movements of the oxygen atoms in the
CuO$_2$ plane, while suggesting that modes involving apical and
chain oxygen are less strongly coupled to the carriers.

\section{Conclusions}\label{sec:conclusion}

In conclusion, the unconventional isotope effects presented here
clearly demonstrate that lattice effects play an important role in
the physics of cuprates. The fact that a substantial OIE on
$\lambda_{ab}$ is observed, even in optimally doped cuprates,
strongly suggests that cuprate HTS are {\em non-adiabatic}
superconductors.  This is in contrast to the novel
high-temperature superconductor MgB$_2$ for which within
experimental error no Boron ($^{10}$B/$^{11}$B) isotope effect on
the magnetic penetration depth
was detected \cite{DiCastro03}.  %
The site-selective OIE studies of $T_c$ and $\lambda_{ab}$
indicate that the phonon modes involving the movements of oxygen
within the superconducting CuO$_2$ planes are essential for the
occurrence of superconductivity in cuprate HTS. This is in
agreement with recent inelastic neutron scattering
\cite{Mcqueeney99,Chung03} and photoemission
\cite{Bogdanov00,Lanzara01} studies, indicating  a strong
interaction between charge carriers and Cu-O bond-stretching-type
of phonons. The generic relation between the OIE on $T_c$ and
$\lambda_{ab}$ demonstrates that, in contrast to conventional
phonon-mediated superconductors, $\lambda_{ab}$ (superfluid
density)  is a key parameter for understanding the role of phonons
in cuprates, in particular local lattice distortions involving
planar oxygen.
The present results rise serious doubts that models, neglecting
lattice degrees of freedom, as proposed for instance in
\cite{Anderson03}, are potential candidates to explain
superconductivity in HTS.

\vspace{0.5cm}

This work was partly performed at the Swiss Muon Source (S$\mu$S),
Paul Scherrer Institute (PSI, Switzerland). The authors are
grateful to A.~Amato U.~Zimmermann, and D.~Herlach for help during
the $\mu$SR measurements, D.~Di~Castro, N.~Garifianov,
D.G.~Eshchenko, H.~Luetkens, T.~Prokscha, and A.~Suter for helpful
discussions and participating in some $\mu$SR experiments,
A.~Bussmann-Holder, E.~Liarokapis, D.~Lampakis, M.~Mali,
K.A.~M\"uller, J.~Roos, T.~Schneider, Z.-X.~Shen, A.~Tatsi, and
G.M.~Zhao for their collaborations and discussions. This work was
supported by the Swiss National Science Foundation and by the NCCR
program \textit{Materials with Novel Electronic Properties}
(MaNEP) sponsored by the Swiss National Science Foundation.


%

\end{document}